\documentclass[aps,prl,twocolumn,showpacs,superscriptaddress]{revtex4-1}

\usepackage{amsmath}
\usepackage{amssymb}
\usepackage{amsfonts}
\usepackage{graphicx}
\usepackage{comment}
\usepackage{xcolor}

\newcommand{\eg}{e.\,g., }
\newcommand{\ie}{i.\,e., }

\begin{document}

\title{Sequence determines degree of knottedness in a coarse-grained
  protein model}

\author{Thomas W\"ust}
\email{twuest@ethz.ch}
\affiliation{Scientific IT Services, ETH Z\"urich, 8092 Z\"urich, Switzerland}
\author{Daniel Reith}
\affiliation{Institut f\"ur Physik, Johannes Gutenberg-Universit\"at Mainz, Staudinger Weg 7, 55099 Mainz, Germany}
\author{Peter Virnau}
\email{virnau@uni-mainz.de}
\affiliation{Institut f\"ur Physik, Johannes Gutenberg-Universit\"at Mainz, Staudinger Weg 7, 55099 Mainz, Germany}


\begin{abstract}
  Knots are abundant in globular homopolymers but rare in globular
  proteins. To shed new light on this long-standing conundrum, we
  study the influence of sequence on the formation of knots in
  proteins under native conditions within the framework of the
  hydrophobic-polar (HP) lattice protein model. By employing large
  scale Wang-Landau simulations combined with suitable Monte Carlo
  trial moves we show that, even though knots are still abundant on
  average, sequence introduces large variability in the degree of
  self-entanglements. Moreover, we are able to design sequences which
  are either almost always or almost never knotted. Our findings serve
  as proof of concept that the introduction of just one additional
  degree of freedom per monomer (in our case sequence) facilitates
  evolution towards a protein universe in which knots are rare.
\end{abstract}


\maketitle

Knots have fascinated physicists, mathematicians and chemists for a
long time. About 140 years ago, Kelvin hypothesized that atoms consist
of knots in the aether \cite{thomson:p_roy_soc_edinb:67}. At first
sight, this beautiful idea is quite appealing as knots are, in a
sense, unique and just like atoms cannot change their type: Without
breaking bonds a simple unknotted ring (a so-called unknot) cannot be,
\eg transformed into a trefoil knot ($3_1$, with three minimal
crossings in a projection onto a plane). But as this aesthetically
pleasing model was finally rejected most of the initial enthusiasm
among natural scientists faded, and knot theory became truly a part of
mathematical sciences. In recent decades, however, the field went
through a renaissance spurred by the discovery of knots in DNA
\cite{liu:j_mol_biol:76,arsuaga:pnas:02} and proteins
\cite{mansfield:nat_struct_mol_biol:94,taylor:nature:00,virnau:j_phys_condens_matter:11}.

Knotted proteins in particular pose a number of challenges which are
not overcome easily and question our understanding of evolution and
folding - especially when we keep in mind that the function of a
protein is determined by its three-dimensional structure. Only eleven
folds are known to be knotted (one of which has been created
artificially) and most of these knots are simple trefoils
\cite{virnau:j_phys_condens_matter:11}. There is also one protein knot
with five crossings which incidentally makes up 1-2\,\% of our brain
protein mass, (pdb-code:2etl) \cite{virnau:plos:06}, and there is even
a knot with six crossings (pdb-code:3bjx)
\cite{boelinger:plos:10}. Indeed, it is difficult to imagine how such
proteins always fold into their knotted native state
\cite{sulkowska:pnas:09}. A number of experiments have shown that
certain knotted proteins can refold to the knotted state upon
degradation \cite{mallam:j_mol_biol:05} and that the process can be
accelerated by chaperons \cite{mallam:nat_chem_biol:12}. From a
topological point of view folding may not always be as difficult as it
appears in the first place though, as even complicated knots (\eg the
$6_1$ knot mentioned above) can be generated from an unknotted state
by a single global movement of a subchain as shown by coarse-grained
folding simulations with G\~o-models \cite{boelinger:plos:10}.

The apparent rarity of knotted proteins is in stark contrast to the
abundance of knots in globular polymers
\cite{mansfield:macromolecules:94,lua:polymer:04,virnau:j_am_chem_soc:05,micheletti:phys_rep:11}. Even
though proteins are not archetypal homopolymers of the bead-spring
type, this discrepancy is nevertheless remarkable. Indeed, there are
several competing (and even complementing) ideas why knots are
rare. Taylor and Lin \cite{taylor:nature:03} pointed out that proteins
should rather be compared to a chain of ``sticky beads'' -- a
visualization of an old idea \cite{grosberg:j_phys:88}: The protein
essentially folds from an unknotted swollen state and remains in an
unknotted (``crumbled'') globular state which results from the initial
collapse. From a structural point of view the emergence of secondary
structure also changes the length-scale at which knots occur and
likely decreases their probability of occurrence. A first systematic
study in this context was undertaken by Lua and Grosberg
\cite{lua:plos:06}; they compared the scaling of subchains between
real proteins and compact lattice loops and found that, statistically,
proteins tend to ``fold back on themselves'' at intermediate scales up
to 40 amino acids which may act as a strong suppressor of knotting. To
which extent this is a result of evolution working towards the
suppression of knots (as they may be adverse to folding or function)
is still largely unknown.

In this letter we focus on how such mechanisms may have evolved in the
first place. Consider a statistical ensemble of (potentially highly
knotted) globular proteins made up of random amino acids with a
certain degree of variability. Natural selection has led to a
``Protein Universe'' \cite{dawkins:96,dill:protein_sci:99}
significantly different from the statistical average of our random
amino acid chains - apparently full of purpose and function and with
little or no knots.

Within the framework of a minimalist protein model, the
hydrophobic-polar (HP) lattice model
\cite{dill:biochemistry:85,lau:macromolecules:89,dill:protein_sci:95},
we show that a single additional degree of freedom per monomer, namely
sequence, may provide an evolutionary pathway which allows proteins to
evolve towards a ``lattice protein universe'' which is almost void of
knots. We are able to design sequences and identify patterns, which
suppress or enhance the formation of knots in our lattice
model. However, due to the coarse-grained nature of the lattice, these
sequences are typically not the same as in real proteins which are
considerably more complex.

In the HP model the protein is represented as a self-avoiding chain of
beads (the amino acid residues) on a regular lattice (here, simple
cubic). There are only two classes of amino acids, hydrophobic (H) and
polar (P) residues. Proteins as opposed to homopolymers have a
hydrophobic core resulting from the tendency of shielding the
hydrophobic side-chains from the polar (aqueous) environment. In the
HP model this hydrophobic force is (implicitly) mimicked by an
attractive interaction $\epsilon$ that acts between non-bonded
neighboring H residues ($\epsilon_{HH} = -1$, $\epsilon_{HP, PP} =
0$). Thus, at low temperatures H residues tend to gather in the
interior of the globular state and form a hydrophobic core while P
residues are located at the outer shell. Despite its limitations
\cite{chan:proteins:00,rios:phys_rev_E:00}, the HP model has been
widely used to describe protein folding qualitatively and to shed new
light onto some of the most puzzling questions in protein science (\eg
Levinthal and blind watchmaker paradox \cite{dill:protein_sci:99},
chaperonin-mediated protein folding \cite{chan:proteins:96},
mutation-induced fold switching \cite{holzgrafe:j_chem_phys:11}, to
mention a few). Thus, it also serves as a good starting point to
address the questions of knottedness (a fundamental, topological
property of proteins) at the level of abstraction of the present
study. While the result of a successful folding process along a
folding funnel is generally assumed to correspond to a free energy
minimum \cite{bryngelson:proteins:95,dill:protein_sci:99}, our
simulations generate conformations in the close vicinity of this
minimum, which are subsequently analyzed with respect to knots
\footnote{Of course, the paradigm that ground-states correspond to
  free energy minima of folding trajectories implicitly assumes that
  folding does not get stuck in kinetic traps which lead to
  topologically frustrated (unknotted) states. The investigation of
  this interesting and challenging topic is, however, beyond the scope
  of this paper.}.

In order to address the problem from a statistical point of view we
need to sample a large ensemble of random protein sequences under
\emph{native} conditions (\ie ground-state like). To make sure that
lattice effects do not bias the statistics, long chains lengths ($N >
100$) are required \cite{boelinger:08}. Together, these requirements
pose a considerable challenge on the computational procedure and,
thus, a similar systematic study has not been carried out for any type
of protein model so far. Even for the very simplified HP model,
estimating the ground-state of a specific HP sequence has only been
possible up to around 100 monomers with state of the art techniques
and computational
power~\cite{hsu:phys_rev_E:03,*hsu:j_chem_phys:03,backofen:constraints:06,zhang:j_chem_phys:07}.

Recently, however, W\"ust and Landau
\cite{wust:phys_rev_lett:09,*wust:j_chem_phys:12} proposed an
efficient Monte Carlo scheme which renders the sampling of
\emph{uncorrelated}, low-energy (\ie ``native like'') structures
feasible even for chain lengths up to $N = 500$. The key of their
procedure is the combination of Wang-Landau (WL) sampling
\cite{wang:phys_rev_lett:01,*wang:phys_rev_E:01} with two
non-traditional Monte Carlo trial moves, namely pull moves
\cite{lesh:recomb:03} and bond-rebridging moves
\cite{deutsch:j_chem_phys:97} which complement each other extremely
well. Their methodology has proven to be very powerful in overcoming
both the energetic and entropic barriers typically encountered when
sampling the complex free energy landscape of dense lattice polymers
and proteins. For details, see \cite{wust:j_chem_phys:12}.

For the topological characterization of protein conformations we need
to compute so called knot invariants which are only unique for closed
curves (mathematically, knots are only well-defined for closed
curves). Thus, for linear polymers and proteins the notion of
knottedness needs to be extended to open chains by choosing a
particular closure which connects the termini in a well-defined manner
(thus closing the loop)
\cite{millett:macromolecules:05,tubiana:prog_theor_phys_suppl:11,millett:biochem_soc_trans:13}. It
is important that the closure itself has no significant influence on
the calculation of the knot invariants. Even though some ambiguity
remains, different closures typically yield similar results from a
statistical point of view
\cite{virnau:j_am_chem_soc:05,virnau:plos:06}. In this paper we use a
rather simple closure which was already applied successfully for the
determination of knots in real proteins: We determine the center of
mass of the polymer and draw two lines through the first and the last
bead. Outside the protein the two lines are connected by a straight
line. From this structure we compute the Alexander polynomial (knot
invariant). The numerical implementation of the entire procedure is
described in great detail in
\cite{virnau:plos:06,virnau:phys_procedia:10}.

\begin{figure}

  \centering

  \includegraphics[width=1.0\columnwidth,clip]{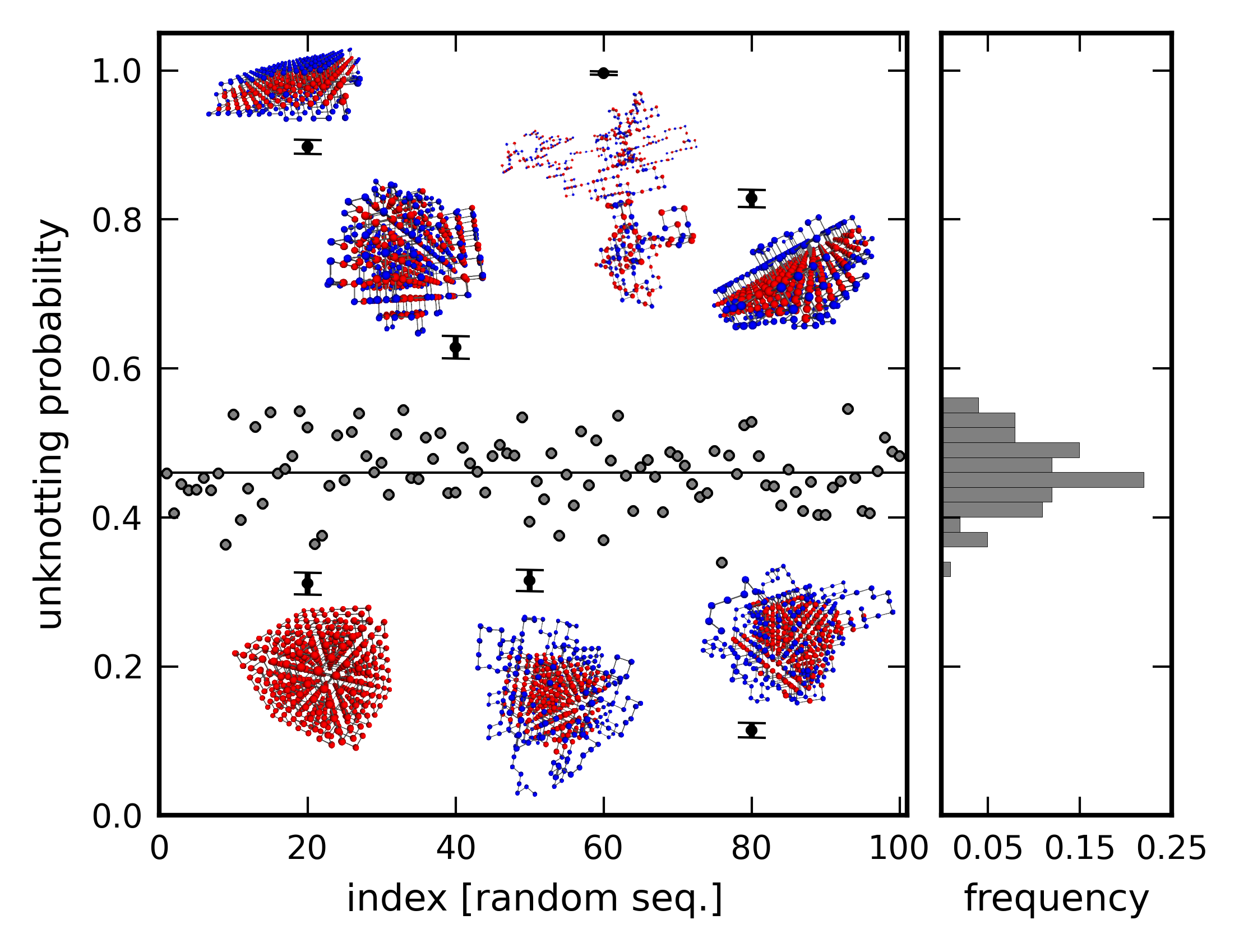}

  \caption{\label{fig:knotting_fluct} \emph{Left panel:} Unknotting
    probabilities of 100 random HP sequences (\emph{gray dots}) and
    selected, designed sequences (\emph{black dots with error bars})
    with $N = 500$ monomers under native conditions; for the latter,
    representative snapshots of native conformations are shown
    too. The designed sequences are: (\emph{upper part, from left to
      right}) seq.~\#1 (see Fig.~\ref{fig:snapshots});
    $\mathrm{PH(PHPHPHHPHPHHPPHP)_{31}HP}$; self-avoiding walk
    ($\mathrm{[HP]_{250}}$);
%
%
    $\mathrm{(P_2H_2)_{12}(P_4H_4)_8(P_2H_2)_{12}(P_4H_4)_8(P_2H_2)_6PH}$
    + same sequence in reverse order; (\emph{lower part, from left to
      right}) homopolymer ($\mathrm{H_{500}}$);
    $\mathrm{(H_{10}P_{10})_{25}}$; seq.~\#2 (see
    Fig.~\ref{fig:snapshots}). Error bars of unknotting probabilities
    for the individual sequences have been estimated by a jackknife
    analysis (because of similarity, only shown for the designed
    sequences). Note that the distribution of points on the x-axis is
    arbitrary. The mean unknotting probability of the 100 random
    sequences is 0.460(5) (\emph{thin horizontal line}). \emph{Right
      panel:} Frequency distribution of unknotting probabilities of
    the 100 random HP sequences.}

\end{figure}

Fig.~\ref{fig:knotting_fluct} shows the unknotting probabilities for
100 random HP sequences and a few designed HP sequences under native
conditions. All chains consist of $N = 500$ monomers with 50\,\% H and
50\,\% P residues (except for the homopolymer with 100\,\% H). This
chain length was chosen such that the homopolymer already exhibit a
significant amount of knotting. We have also studied shorter chains
(down to $N = 100$) and obtained qualitatively similar results even
though the overall probability to find a knot for shorter chain
lengths is, of course, correspondingly smaller. In each simulation the
sequence of a chain is fixed and does not vary. Thus, we investigate
an ensemble of sequences to show how the introduction of this
additional degree of freedom per monomer may affect an evolutionary
system.

It is worth noting that the HP model exhibits a rather large
ground-state degeneracy, which could be reduced somewhat by adding
additional interactions between H and P monomers. The degeneracy of
the ground-state and the additional states in its vicinity allowed us,
however, to determine a "likelihood" of knottedness for a given HP
sequence as follows: First, a pre-WL run was performed to obtain an
estimate of its ground-state energy. Then, a subsequent production WL
simulation, restricted to the lowest 20\,\% of the entire energy
range, consecutively sampled conformations within 5\,\% of the
ground-state energy. (This threshold was set heuristically but other
values $<10$\,\% gave similar results). Between the sampling of any
two conformations the random walker must always perform a full round
trip through the specified energy range in order to reduce possible
structural correlations. Multiple, independent production WL
simulations were run simultaneously to speed up the sampling and
further increase the structural diversity of sampled
conformations. Eventually, 1000 conformations were randomly selected
among the entire sample and their knottedness analyzed. The unknotting
probability, as displayed in Fig.~\ref{fig:knotting_fluct}, is then
defined as the number of unknotted conformations divided by 1000
\footnote{Note that for an estimate of the free energy minimum
  knottedness, \ie at zero temperature, an accurate determination of
  the density of states is irrelevant and WL sampling has merely been
  used as a powerful Monte Carlo driver to sample statistically
  uncorrelated, low-energy conformations.}.
The total sampling time for the whole study amounted to more than
three million CPU hours (AMD Opteron 6272, 2.1\,GHz).

To interpret this probability we can imagine that a HP lattice polymer
with a given sequence represents a large number of possible proteins
with the same or a very similar sequence of hydrophobic and polar
amino acids. Small changes in interactions (representing, \eg amino
acids with a slightly different degree of hydrophobicity) or slightly
different sequences will lead to similar knotting behavior. To check
this assumption we have performed additional simulations in which we
randomly mutated monomers (while keeping the ratio of H and P residues
constant) and have confirmed that the likelihood of containing knots
is indeed similar for mutation fractions up to 4\,\%. Hence, the
unknotting probability can be interpreted as an estimate for the
unknotted fraction of conformational space of proteins represented by
this sequence.

Fig.~\ref{fig:knotting_fluct} clearly illustrates the strong
dependence of the degree of knottedness on the particular
sequence. Even for the random sequences the unknotting probability
fluctuates between 0.3 and 0.6. Despite this large variability in the
tendency of individual sequences to form knots, on average
heteropolymers are almost as knotted as globular homopolymers. Most
remarkably, however, is the fact that it is possible to design HP
sequences (notably, with the same ratio of H and P) featuring almost
no knots or being almost fully knotted.

\begin{figure}

  \centering

  \includegraphics[width=1.0\columnwidth,clip]{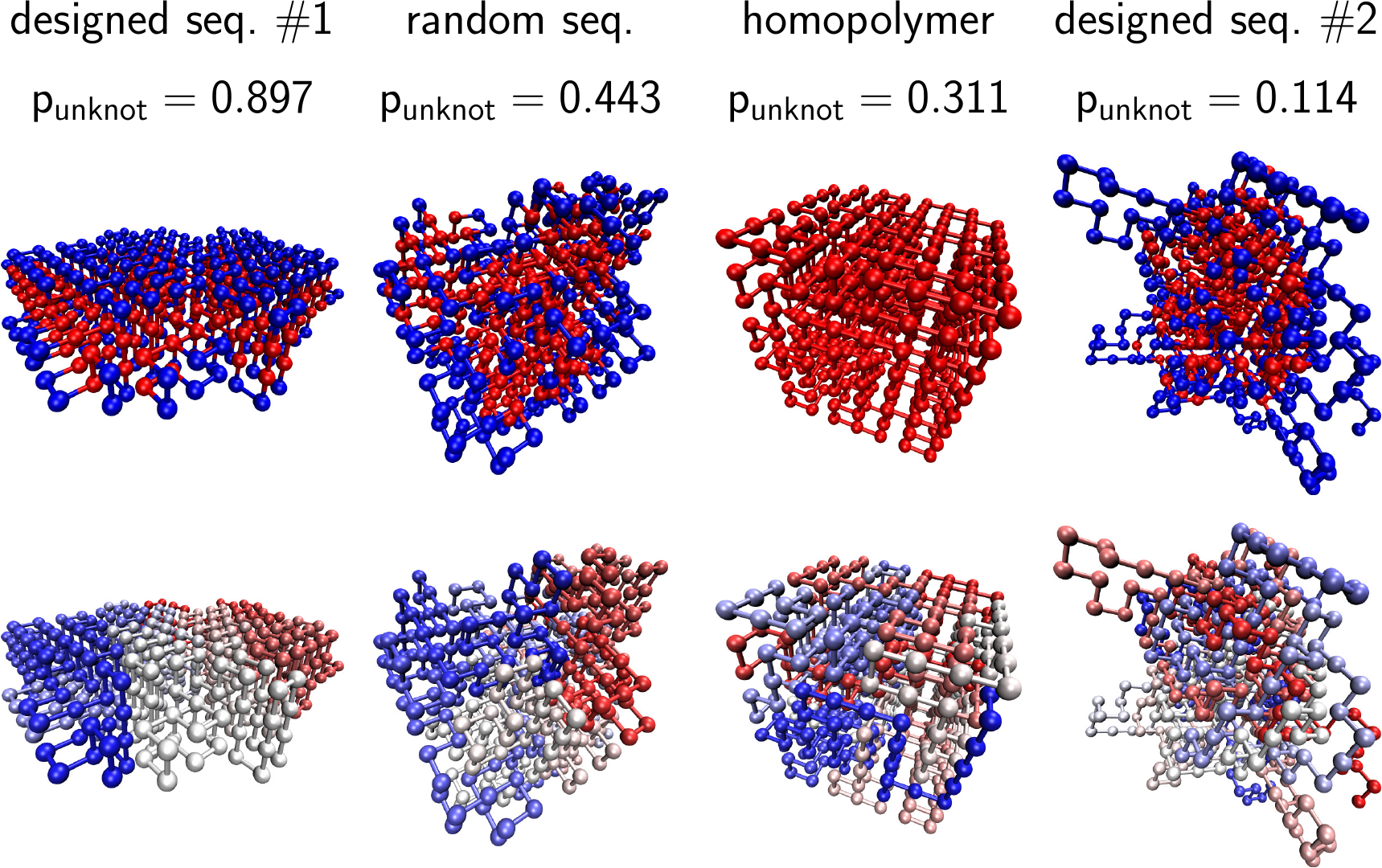}

  \caption{\label{fig:snapshots} Snapshots of representative native
    structures. \emph{From left to right:} Designed HP sequence
    featuring almost no knots ($\mathrm{[HHPP]_{125}}$); random HP
    sequence with a modest degree of knottedness; homopolymer (an
    exact ground-state structure); highly knotted designed HP sequence
    ($\mathrm{[P_{10}(HP)_7H_{10}(HP)_8]_5[(PH)_8H_{10}(PH)_7P_{10}]_5}$). \emph{Upper
      structures:} Monomer type coloring with hydrophobic monomers and
    polar monomers shown in red and blue, respectively. \emph{Lower
      structures:} Monomer index coloring, with colors gradually
    changing from blue (monomers at the beginning of the sequence),
    over white (monomers in the center of the sequence) to red
    (monomers at the end of the sequence). $p_{\text{unknot}}$ denotes
    the unknotting probability of the corresponding sequence.}

\end{figure}

Fig.~\ref{fig:snapshots} shows snapshots of typical native state like
structures for a random HP sequence, two designed HP sequences, and
the homopolymer. Lattice homopolymers close to the native state are
cubic, but have little local order: Inside the cube the chain goes
back and forth leading to a rather large degree of knottedness. A
typical random HP sequence already has a pronounced hydrophobic core
and tends to be a bit more ordered at the local scale: Beads which are
only a few monomers apart tend to occupy the same region in
space. This leads to a small decrease in the overall knotting
probability, but by no means explains the large discrepancy between
proteins and homopolymers. The two designed sequences, which are
extreme examples with respect to the variability in the tendency to
form knots, exhibit very pronounced and distinct features. The pattern
of alternating HH and PP segments in the designed sequence \#1, which
is almost always unknotted, induces a very regular, slab-shaped,
native structure with H monomers filling the interior of the slab and
P monomers occupying its border. This compact structure results from a
distinct local threading of the sequence which disfavors
entanglements. Alternating sequences of $\mathrm{H_4P_4}$ and
$\mathrm{H_2P_2}$ segments have a similar effect (cf.\ snapshot in
Fig.~\ref{fig:knotting_fluct}), but structures tend to be ellipsoidal
rather than flat. Another, almost trivial motive are simple
alternating H and P monomers, which form a swollen coil structure akin
to a self-avoiding walk (cf.\ snapshot in
Fig.~\ref{fig:knotting_fluct}). In contrast, a pattern which highly
favors the formation of knots is presented in the designed sequence
\#2. It consists of long contiguous segments of H and P residues
separated by segments of repeating (HP) motives. This sequence forces
the protein to fold back through extended loops in order to optimize
the number of non-bonded HH interactions and there is almost no local
order inside the hydrophobic core. Both features foster entanglements
and knots. Again, we need to stress that there is no one-to-one
correspondence between motives which enhance or suppress knots in
lattice proteins and real proteins, which are considerably more
complex. Indeed, for the latter such motives have not even been
identified.

Despite the variation in the degree of knottedness, all native like
structures of HP sequences exhibit a more or less pronounced
hydrophobic core. Thus, the formation of a hydrophobic core in itself
cannot be considered as a precursor of suppression of knots in
proteins. However, the local structure (order) among residues within a
sequence strongly influences knotting as manifested by the index
coloring scheme of corresponding structures (see lower row of
snapshots in Fig.~\ref{fig:snapshots}). Whereas in the designed
sequence \#1 nearby monomers are strongly localized and form a
precursor of secondary structure, in the designed sequence \#2 they
tend to spread out far and in uncorrelated directions. In real
proteins individual elements of secondary structure have the tendency
to fold back onto themselves which, in turn, introduces locality and
suppresses knots \cite{lua:plos:06}. It is remarkable that the HP
sequences studied here show the same relationship between knottedness
and local structure despite the simplicity of the underlying protein
model.

\begin{figure}

  \centering

  \includegraphics[width=1.0\columnwidth,clip]{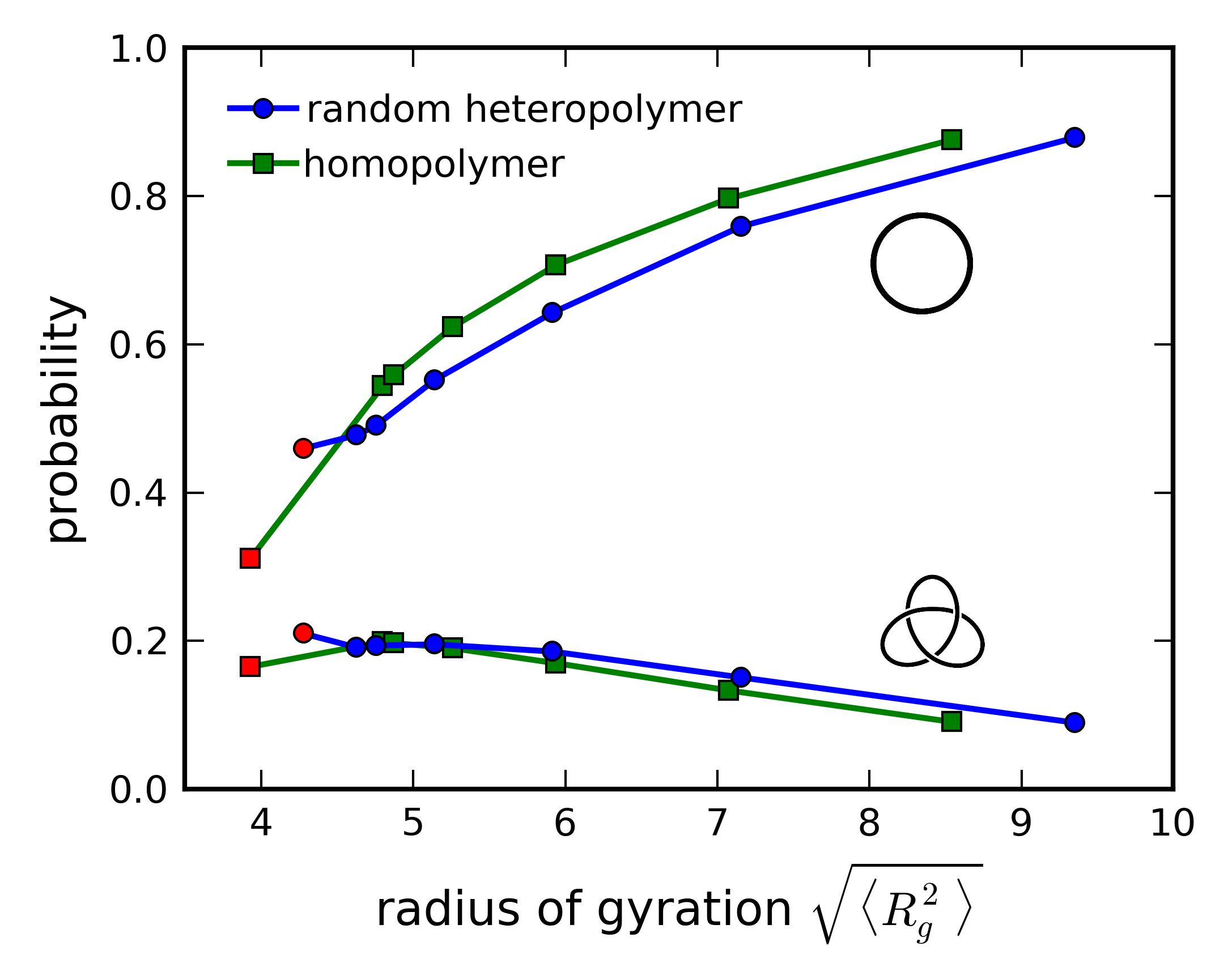}

  \caption{\label{fig:knotting_mean} Average probabilities of finding
    unknots (\emph{upper curves}) and trefoils (\emph{lower curves}),
    respectively, in homopolymers (100\,\% H) and random
    heteropolymers (50\,\% H, 50\,\% P) with $N = 500$ as a function
    of the root mean squared radius of gyration, $\sqrt{\langle R_g^2
      \rangle}$. The red symbols denote corresponding probabilities
    under native condition (ground-state like). Error bars have been
    calculated by averaging over independent runs; they do not exceed
    symbol size and are, thus, not shown.}

\end{figure}

Finally, we compare the average knotting probability of random hetero-
and homopolymers as a function of solvent quality (\ie temperature in
our model) ranging from ground-state like structures, in which knots
tend to spread over the whole structure, to the denatured case, in
which they are weakly localized (not shown here). To make a fair
comparison, we plot the probability of observing an unknotted
structure (or a trefoil knot) as a function of the radius of
gyration. To be able to define an average knotting probability for
random heteropolymers, we have again averaged over our 100 random HP
sequences; Metropolis Monte Carlo sampling (using the same move sets
as described above) has been employed to obtain correctly weighted
estimates at finite temperatures. Fig.~\ref{fig:knotting_mean} shows
that the probability of finding unknots or trefoil knots in
heteropolymers (averaged over random sequences) is quite similar to
the one for homopolymers at comparable densities. However, at high
densities (low temperatures) the unknotting probability of
heteropolymers clearly deviates from the decreasing trend observed in
homopolymers.

In this study we have been able to demonstrate quantitatively that
sequence strongly influences (or even determines) the degree of
knottedness under native conditions. Within the framework of the
minimalist HP protein model and large scale Monte Carlo simulations,
we have determined probabilities of knotting for random HP sequences
as well as homopolymers with 500 residues. The introduction of
sequence leads to a large variability in the self-entanglements of
heteropolymers even though on average they are almost as knotted as
globular homopolymers of comparable density. We have also been able to
design sequences which fold into either highly knotted or almost
knot-less structures. While we demonstrate that a variation of
sequence leads to a variation of self-entanglements and knots it is
likely that variability in other interactions may have similar
effects. This shows in principle that the introduction of a single
additional degree of freedom per monomer, in our case sequence, may
already suffice to facilitate evolution towards a largely unknotted
``Protein Universe''. In a sense, proteins are not an equilibrium
ensemble of (knotted) random heteropolymers and should as such not be
compared to an equivalent ensemble of homopolymers, but instead live
in a very specific conformational subspace in which knots are rare.

We thank D.~B\"olinger and H.-P.~Hsu for stimulating discussions. This
work received partial financial support from the Deutsche
Forschungsgemeinschaft (SFB 625-A17).

\bibliography{paper}

\begin{thebibliography}{45}%
\makeatletter
\providecommand \@ifxundefined [1]{%
 \@ifx{#1\undefined}
}%
\providecommand \@ifnum [1]{%
 \ifnum #1\expandafter \@firstoftwo
 \else \expandafter \@secondoftwo
 \fi
}%
\providecommand \@ifx [1]{%
 \ifx #1\expandafter \@firstoftwo
 \else \expandafter \@secondoftwo
 \fi
}%
\providecommand \natexlab [1]{#1}%
\providecommand \enquote  [1]{``#1''}%
\providecommand \bibnamefont  [1]{#1}%
\providecommand \bibfnamefont [1]{#1}%
\providecommand \citenamefont [1]{#1}%
\providecommand \href@noop [0]{\@secondoftwo}%
\providecommand \href [0]{\begingroup \@sanitize@url \@href}%
\providecommand \@href[1]{\@@startlink{#1}\@@href}%
\providecommand \@@href[1]{\endgroup#1\@@endlink}%
\providecommand \@sanitize@url [0]{\catcode `\\12\catcode `\$12\catcode
  `\&12\catcode `\#12\catcode `\^12\catcode `\_12\catcode `\%12\relax}%
\providecommand \@@startlink[1]{}%
\providecommand \@@endlink[0]{}%
\providecommand \url  [0]{\begingroup\@sanitize@url \@url }%
\providecommand \@url [1]{\endgroup\@href {#1}{\urlprefix }}%
\providecommand \urlprefix  [0]{URL }%
\providecommand \Eprint [0]{\href }%
\providecommand \doibase [0]{http://dx.doi.org/}%
\providecommand \selectlanguage [0]{\@gobble}%
\providecommand \bibinfo  [0]{\@secondoftwo}%
\providecommand \bibfield  [0]{\@secondoftwo}%
\providecommand \translation [1]{[#1]}%
\providecommand \BibitemOpen [0]{}%
\providecommand \bibitemStop [0]{}%
\providecommand \bibitemNoStop [0]{.\EOS\space}%
\providecommand \EOS [0]{\spacefactor3000\relax}%
\providecommand \BibitemShut  [1]{\csname bibitem#1\endcsname}%
\let\auto@bib@innerbib\@empty
\bibitem [{\citenamefont {Thomson}(1867)}]{thomson:p_roy_soc_edinb:67}%
  \BibitemOpen
  \bibfield  {author} {\bibinfo {author} {\bibfnamefont {W.}~\bibnamefont
  {Thomson}},\ }\href@noop {} {\bibfield  {journal} {\bibinfo  {journal} {P.\
  Roy.\ Soc.\ Edinb.}\ }\textbf {\bibinfo {volume} {VI}},\ \bibinfo {pages}
  {94} (\bibinfo {year} {1867})}\BibitemShut {NoStop}%
\bibitem [{\citenamefont {Liu}\ \emph {et~al.}(1976)\citenamefont {Liu},
  \citenamefont {Depew},\ and\ \citenamefont {Wang}}]{liu:j_mol_biol:76}%
  \BibitemOpen
  \bibfield  {author} {\bibinfo {author} {\bibfnamefont {L.~F.}\ \bibnamefont
  {Liu}}, \bibinfo {author} {\bibfnamefont {R.~E.}\ \bibnamefont {Depew}}, \
  and\ \bibinfo {author} {\bibfnamefont {J.~C.}\ \bibnamefont {Wang}},\
  }\href@noop {} {\bibfield  {journal} {\bibinfo  {journal} {J.\ Mol.\ Biol.}\
  }\textbf {\bibinfo {volume} {106}},\ \bibinfo {pages} {439} (\bibinfo {year}
  {1976})}\BibitemShut {NoStop}%
\bibitem [{\citenamefont {Arsuaga}\ \emph {et~al.}(2002)\citenamefont
  {Arsuaga}, \citenamefont {V{\'a}zquez}, \citenamefont {Trigueros},
  \citenamefont {Sumners},\ and\ \citenamefont {Roca}}]{arsuaga:pnas:02}%
  \BibitemOpen
  \bibfield  {author} {\bibinfo {author} {\bibfnamefont {J.}~\bibnamefont
  {Arsuaga}}, \bibinfo {author} {\bibfnamefont {M.}~\bibnamefont
  {V{\'a}zquez}}, \bibinfo {author} {\bibfnamefont {S.}~\bibnamefont
  {Trigueros}}, \bibinfo {author} {\bibfnamefont {D.~W.}\ \bibnamefont
  {Sumners}}, \ and\ \bibinfo {author} {\bibfnamefont {J.}~\bibnamefont
  {Roca}},\ }\href@noop {} {\bibfield  {journal} {\bibinfo  {journal} {Proc.\
  Natl.\ Acad.\ Sci.\ USA}\ }\textbf {\bibinfo {volume} {99}},\ \bibinfo
  {pages} {5373} (\bibinfo {year} {2002})}\BibitemShut {NoStop}%
\bibitem [{\citenamefont
  {Mansfield}(1994{\natexlab{a}})}]{mansfield:nat_struct_mol_biol:94}%
  \BibitemOpen
  \bibfield  {author} {\bibinfo {author} {\bibfnamefont {M.~L.}\ \bibnamefont
  {Mansfield}},\ }\href@noop {} {\bibfield  {journal} {\bibinfo  {journal}
  {Nat.\ Struct.\ Mol.\ Biol.}\ }\textbf {\bibinfo {volume} {1}},\ \bibinfo
  {pages} {213} (\bibinfo {year} {1994}{\natexlab{a}})}\BibitemShut {NoStop}%
\bibitem [{\citenamefont {Taylor}(2000)}]{taylor:nature:00}%
  \BibitemOpen
  \bibfield  {author} {\bibinfo {author} {\bibfnamefont {W.~R.}\ \bibnamefont
  {Taylor}},\ }\href@noop {} {\bibfield  {journal} {\bibinfo  {journal}
  {Nature}\ }\textbf {\bibinfo {volume} {406}},\ \bibinfo {pages} {916}
  (\bibinfo {year} {2000})}\BibitemShut {NoStop}%
\bibitem [{\citenamefont {Virnau}\ \emph {et~al.}(2011)\citenamefont {Virnau},
  \citenamefont {Mallam},\ and\ \citenamefont
  {Jackson}}]{virnau:j_phys_condens_matter:11}%
  \BibitemOpen
  \bibfield  {author} {\bibinfo {author} {\bibfnamefont {P.}~\bibnamefont
  {Virnau}}, \bibinfo {author} {\bibfnamefont {A.}~\bibnamefont {Mallam}}, \
  and\ \bibinfo {author} {\bibfnamefont {S.}~\bibnamefont {Jackson}},\
  }\href@noop {} {\bibfield  {journal} {\bibinfo  {journal} {J.\ Phys.:
  Condens.\ Matter}\ }\textbf {\bibinfo {volume} {23}},\ \bibinfo {pages}
  {033101} (\bibinfo {year} {2011})}\BibitemShut {NoStop}%
\bibitem [{\citenamefont {Virnau}\ \emph {et~al.}(2006)\citenamefont {Virnau},
  \citenamefont {Mirny},\ and\ \citenamefont {Kardar}}]{virnau:plos:06}%
  \BibitemOpen
  \bibfield  {author} {\bibinfo {author} {\bibfnamefont {P.}~\bibnamefont
  {Virnau}}, \bibinfo {author} {\bibfnamefont {L.~A.}\ \bibnamefont {Mirny}}, \
  and\ \bibinfo {author} {\bibfnamefont {M.}~\bibnamefont {Kardar}},\
  }\href@noop {} {\bibfield  {journal} {\bibinfo  {journal} {PLoS Comput.\
  Biol.}\ }\textbf {\bibinfo {volume} {2}},\ \bibinfo {pages} {e122} (\bibinfo
  {year} {2006})}\BibitemShut {NoStop}%
\bibitem [{\citenamefont {B{\"o}linger}\ \emph {et~al.}(2010)\citenamefont
  {B{\"o}linger}, \citenamefont {Su{\l}kowska}, \citenamefont {Hsu},
  \citenamefont {Mirny}, \citenamefont {Kardar}, \citenamefont {Onuchic},\ and\
  \citenamefont {Virnau}}]{boelinger:plos:10}%
  \BibitemOpen
  \bibfield  {author} {\bibinfo {author} {\bibfnamefont {D.}~\bibnamefont
  {B{\"o}linger}}, \bibinfo {author} {\bibfnamefont {J.~I.}\ \bibnamefont
  {Su{\l}kowska}}, \bibinfo {author} {\bibfnamefont {H.-P.}\ \bibnamefont
  {Hsu}}, \bibinfo {author} {\bibfnamefont {L.~A.}\ \bibnamefont {Mirny}},
  \bibinfo {author} {\bibfnamefont {M.}~\bibnamefont {Kardar}}, \bibinfo
  {author} {\bibfnamefont {J.~N.}\ \bibnamefont {Onuchic}}, \ and\ \bibinfo
  {author} {\bibfnamefont {P.}~\bibnamefont {Virnau}},\ }\href@noop {}
  {\bibfield  {journal} {\bibinfo  {journal} {PLoS Comput.\ Biol.}\ }\textbf
  {\bibinfo {volume} {6}},\ \bibinfo {pages} {e1000731} (\bibinfo {year}
  {2010})}\BibitemShut {NoStop}%
\bibitem [{\citenamefont {Su{\l}kowska}\ \emph {et~al.}(2009)\citenamefont
  {Su{\l}kowska}, \citenamefont {Su{\l}kowski},\ and\ \citenamefont
  {Onuchic}}]{sulkowska:pnas:09}%
  \BibitemOpen
  \bibfield  {author} {\bibinfo {author} {\bibfnamefont {J.~I.}\ \bibnamefont
  {Su{\l}kowska}}, \bibinfo {author} {\bibfnamefont {P.}~\bibnamefont
  {Su{\l}kowski}}, \ and\ \bibinfo {author} {\bibfnamefont {J.}~\bibnamefont
  {Onuchic}},\ }\href@noop {} {\bibfield  {journal} {\bibinfo  {journal}
  {Proc.\ Natl.\ Acad.\ Sci.\ USA}\ }\textbf {\bibinfo {volume} {106}},\
  \bibinfo {pages} {3119} (\bibinfo {year} {2009})}\BibitemShut {NoStop}%
\bibitem [{\citenamefont {Mallam}\ and\ \citenamefont
  {Jackson}(2005)}]{mallam:j_mol_biol:05}%
  \BibitemOpen
  \bibfield  {author} {\bibinfo {author} {\bibfnamefont {A.~L.}\ \bibnamefont
  {Mallam}}\ and\ \bibinfo {author} {\bibfnamefont {S.~E.}\ \bibnamefont
  {Jackson}},\ }\href@noop {} {\bibfield  {journal} {\bibinfo  {journal} {J.\
  Mol.\ Biol.}\ }\textbf {\bibinfo {volume} {346}},\ \bibinfo {pages} {1409}
  (\bibinfo {year} {2005})}\BibitemShut {NoStop}%
\bibitem [{\citenamefont {Mallam}\ and\ \citenamefont
  {Jackson}(2012)}]{mallam:nat_chem_biol:12}%
  \BibitemOpen
  \bibfield  {author} {\bibinfo {author} {\bibfnamefont {A.~L.}\ \bibnamefont
  {Mallam}}\ and\ \bibinfo {author} {\bibfnamefont {S.~E.}\ \bibnamefont
  {Jackson}},\ }\href@noop {} {\bibfield  {journal} {\bibinfo  {journal} {Nat.\
  Chem.\ Biol.}\ }\textbf {\bibinfo {volume} {8}},\ \bibinfo {pages} {147}
  (\bibinfo {year} {2012})}\BibitemShut {NoStop}%
\bibitem [{\citenamefont
  {Mansfield}(1994{\natexlab{b}})}]{mansfield:macromolecules:94}%
  \BibitemOpen
  \bibfield  {author} {\bibinfo {author} {\bibfnamefont {M.~L.}\ \bibnamefont
  {Mansfield}},\ }\href@noop {} {\bibfield  {journal} {\bibinfo  {journal}
  {Macromolecules}\ }\textbf {\bibinfo {volume} {27}},\ \bibinfo {pages} {5924}
  (\bibinfo {year} {1994}{\natexlab{b}})}\BibitemShut {NoStop}%
\bibitem [{\citenamefont {Lua}\ \emph {et~al.}(2004)\citenamefont {Lua},
  \citenamefont {Borovinskiy},\ and\ \citenamefont
  {Grosberg}}]{lua:polymer:04}%
  \BibitemOpen
  \bibfield  {author} {\bibinfo {author} {\bibfnamefont {R.}~\bibnamefont
  {Lua}}, \bibinfo {author} {\bibfnamefont {A.~L.}\ \bibnamefont
  {Borovinskiy}}, \ and\ \bibinfo {author} {\bibfnamefont {A.~Y.}\ \bibnamefont
  {Grosberg}},\ }\href@noop {} {\bibfield  {journal} {\bibinfo  {journal}
  {Polymer}\ }\textbf {\bibinfo {volume} {45}},\ \bibinfo {pages} {717}
  (\bibinfo {year} {2004})}\BibitemShut {NoStop}%
\bibitem [{\citenamefont {Virnau}\ \emph {et~al.}(2005)\citenamefont {Virnau},
  \citenamefont {Kantor},\ and\ \citenamefont
  {Kardar}}]{virnau:j_am_chem_soc:05}%
  \BibitemOpen
  \bibfield  {author} {\bibinfo {author} {\bibfnamefont {P.}~\bibnamefont
  {Virnau}}, \bibinfo {author} {\bibfnamefont {Y.}~\bibnamefont {Kantor}}, \
  and\ \bibinfo {author} {\bibfnamefont {M.}~\bibnamefont {Kardar}},\
  }\href@noop {} {\bibfield  {journal} {\bibinfo  {journal} {J.\ Am.\ Chem.\
  Soc.}\ }\textbf {\bibinfo {volume} {127}},\ \bibinfo {pages} {15102}
  (\bibinfo {year} {2005})}\BibitemShut {NoStop}%
\bibitem [{\citenamefont {Micheletti}\ \emph {et~al.}(2011)\citenamefont
  {Micheletti}, \citenamefont {Marenduzzo},\ and\ \citenamefont
  {Orlandini}}]{micheletti:phys_rep:11}%
  \BibitemOpen
  \bibfield  {author} {\bibinfo {author} {\bibfnamefont {C.}~\bibnamefont
  {Micheletti}}, \bibinfo {author} {\bibfnamefont {D.}~\bibnamefont
  {Marenduzzo}}, \ and\ \bibinfo {author} {\bibfnamefont {E.}~\bibnamefont
  {Orlandini}},\ }\href@noop {} {\bibfield  {journal} {\bibinfo  {journal}
  {Phys.\ Rep.}\ }\textbf {\bibinfo {volume} {504}},\ \bibinfo {pages} {1}
  (\bibinfo {year} {2011})}\BibitemShut {NoStop}%
\bibitem [{\citenamefont {Taylor}\ and\ \citenamefont
  {Lin}(2003)}]{taylor:nature:03}%
  \BibitemOpen
  \bibfield  {author} {\bibinfo {author} {\bibfnamefont {W.~R.}\ \bibnamefont
  {Taylor}}\ and\ \bibinfo {author} {\bibfnamefont {K.}~\bibnamefont {Lin}},\
  }\href@noop {} {\bibfield  {journal} {\bibinfo  {journal} {Nature}\ }\textbf
  {\bibinfo {volume} {421}},\ \bibinfo {pages} {25} (\bibinfo {year}
  {2003})}\BibitemShut {NoStop}%
\bibitem [{\citenamefont {Grosberg}\ \emph {et~al.}(1988)\citenamefont
  {Grosberg}, \citenamefont {Nechaev},\ and\ \citenamefont
  {Shakhnovich}}]{grosberg:j_phys:88}%
  \BibitemOpen
  \bibfield  {author} {\bibinfo {author} {\bibfnamefont {A.~Y.}\ \bibnamefont
  {Grosberg}}, \bibinfo {author} {\bibfnamefont {S.~K.}\ \bibnamefont
  {Nechaev}}, \ and\ \bibinfo {author} {\bibfnamefont {E.~I.}\ \bibnamefont
  {Shakhnovich}},\ }\href@noop {} {\bibfield  {journal} {\bibinfo  {journal}
  {J.\ Phys.}\ }\textbf {\bibinfo {volume} {49}},\ \bibinfo {pages} {2095}
  (\bibinfo {year} {1988})}\BibitemShut {NoStop}%
\bibitem [{\citenamefont {Lua}\ and\ \citenamefont
  {Grosberg}(2006)}]{lua:plos:06}%
  \BibitemOpen
  \bibfield  {author} {\bibinfo {author} {\bibfnamefont {R.~C.}\ \bibnamefont
  {Lua}}\ and\ \bibinfo {author} {\bibfnamefont {A.~Y.}\ \bibnamefont
  {Grosberg}},\ }\href@noop {} {\bibfield  {journal} {\bibinfo  {journal} {PLoS
  Comput.\ Biol.}\ }\textbf {\bibinfo {volume} {2}},\ \bibinfo {pages} {e45}
  (\bibinfo {year} {2006})}\BibitemShut {NoStop}%
\bibitem [{\citenamefont {Dawkins}(1996)}]{dawkins:96}%
  \BibitemOpen
  \bibfield  {author} {\bibinfo {author} {\bibfnamefont {R.}~\bibnamefont
  {Dawkins}},\ }\href@noop {} {\emph {\bibinfo {title} {The Blind Watchmaker:
  Why the Evidence of Evolution Reveals a Universe without Design}}}\ (\bibinfo
   {publisher} {Norton},\ \bibinfo {address} {New York},\ \bibinfo {year}
  {1996})\BibitemShut {NoStop}%
\bibitem [{\citenamefont {Dill}(1999)}]{dill:protein_sci:99}%
  \BibitemOpen
  \bibfield  {author} {\bibinfo {author} {\bibfnamefont {K.~A.}\ \bibnamefont
  {Dill}},\ }\href@noop {} {\bibfield  {journal} {\bibinfo  {journal} {Protein
  Sci.}\ }\textbf {\bibinfo {volume} {8}},\ \bibinfo {pages} {1166} (\bibinfo
  {year} {1999})}\BibitemShut {NoStop}%
\bibitem [{\citenamefont {Dill}(1985)}]{dill:biochemistry:85}%
  \BibitemOpen
  \bibfield  {author} {\bibinfo {author} {\bibfnamefont {K.~A.}\ \bibnamefont
  {Dill}},\ }\href@noop {} {\bibfield  {journal} {\bibinfo  {journal}
  {Biochemistry}\ }\textbf {\bibinfo {volume} {24}},\ \bibinfo {pages} {1501}
  (\bibinfo {year} {1985})}\BibitemShut {NoStop}%
\bibitem [{\citenamefont {Lau}\ and\ \citenamefont
  {Dill}(1989)}]{lau:macromolecules:89}%
  \BibitemOpen
  \bibfield  {author} {\bibinfo {author} {\bibfnamefont {K.~F.}\ \bibnamefont
  {Lau}}\ and\ \bibinfo {author} {\bibfnamefont {K.~A.}\ \bibnamefont {Dill}},\
  }\href@noop {} {\bibfield  {journal} {\bibinfo  {journal} {Macromolecules}\
  }\textbf {\bibinfo {volume} {22}},\ \bibinfo {pages} {3986} (\bibinfo {year}
  {1989})}\BibitemShut {NoStop}%
\bibitem [{\citenamefont {Dill}\ \emph {et~al.}(1995)\citenamefont {Dill},
  \citenamefont {Bromberg}, \citenamefont {Yue}, \citenamefont {Fiebig},
  \citenamefont {Yee}, \citenamefont {Thomas},\ and\ \citenamefont
  {Chan}}]{dill:protein_sci:95}%
  \BibitemOpen
  \bibfield  {author} {\bibinfo {author} {\bibfnamefont {K.~A.}\ \bibnamefont
  {Dill}}, \bibinfo {author} {\bibfnamefont {S.}~\bibnamefont {Bromberg}},
  \bibinfo {author} {\bibfnamefont {K.}~\bibnamefont {Yue}}, \bibinfo {author}
  {\bibfnamefont {K.~M.}\ \bibnamefont {Fiebig}}, \bibinfo {author}
  {\bibfnamefont {D.~P.}\ \bibnamefont {Yee}}, \bibinfo {author} {\bibfnamefont
  {P.~D.}\ \bibnamefont {Thomas}}, \ and\ \bibinfo {author} {\bibfnamefont
  {H.~S.}\ \bibnamefont {Chan}},\ }\href@noop {} {\bibfield  {journal}
  {\bibinfo  {journal} {Protein Sci.}\ }\textbf {\bibinfo {volume} {4}},\
  \bibinfo {pages} {561} (\bibinfo {year} {1995})}\BibitemShut {NoStop}%
\bibitem [{\citenamefont {Chan}(2000)}]{chan:proteins:00}%
  \BibitemOpen
  \bibfield  {author} {\bibinfo {author} {\bibfnamefont {H.~S.}\ \bibnamefont
  {Chan}},\ }\href@noop {} {\bibfield  {journal} {\bibinfo  {journal}
  {Proteins}\ }\textbf {\bibinfo {volume} {40}},\ \bibinfo {pages} {543}
  (\bibinfo {year} {2000})}\BibitemShut {NoStop}%
\bibitem [{\citenamefont {{De Los Rios}}\ and\ \citenamefont
  {Caldarelli}(2000)}]{rios:phys_rev_E:00}%
  \BibitemOpen
  \bibfield  {author} {\bibinfo {author} {\bibfnamefont {P.}~\bibnamefont {{De
  Los Rios}}}\ and\ \bibinfo {author} {\bibfnamefont {G.}~\bibnamefont
  {Caldarelli}},\ }\href@noop {} {\bibfield  {journal} {\bibinfo  {journal}
  {Phys.\ Rev.\ E}\ }\textbf {\bibinfo {volume} {62}},\ \bibinfo {pages} {8449}
  (\bibinfo {year} {2000})}\BibitemShut {NoStop}%
\bibitem [{\citenamefont {Chan}\ and\ \citenamefont
  {Dill}(1996)}]{chan:proteins:96}%
  \BibitemOpen
  \bibfield  {author} {\bibinfo {author} {\bibfnamefont {H.~S.}\ \bibnamefont
  {Chan}}\ and\ \bibinfo {author} {\bibfnamefont {K.~A.}\ \bibnamefont
  {Dill}},\ }\href@noop {} {\bibfield  {journal} {\bibinfo  {journal}
  {Proteins}\ }\textbf {\bibinfo {volume} {24}},\ \bibinfo {pages} {345}
  (\bibinfo {year} {1996})}\BibitemShut {NoStop}%
\bibitem [{\citenamefont {Holzgr\"afe}\ \emph {et~al.}(2011)\citenamefont
  {Holzgr\"afe}, \citenamefont {Irb\"ack},\ and\ \citenamefont
  {Troein}}]{holzgrafe:j_chem_phys:11}%
  \BibitemOpen
  \bibfield  {author} {\bibinfo {author} {\bibfnamefont {C.}~\bibnamefont
  {Holzgr\"afe}}, \bibinfo {author} {\bibfnamefont {A.}~\bibnamefont
  {Irb\"ack}}, \ and\ \bibinfo {author} {\bibfnamefont {C.}~\bibnamefont
  {Troein}},\ }\href@noop {} {\bibfield  {journal} {\bibinfo  {journal} {J.\
  Chem.\ Phys.}\ }\textbf {\bibinfo {volume} {135}},\ \bibinfo {pages} {195101}
  (\bibinfo {year} {2011})}\BibitemShut {NoStop}%
\bibitem [{\citenamefont {Bryngelson}\ \emph {et~al.}(1995)\citenamefont
  {Bryngelson}, \citenamefont {Onuchic}, \citenamefont {Socci},\ and\
  \citenamefont {Wolynes}}]{bryngelson:proteins:95}%
  \BibitemOpen
  \bibfield  {author} {\bibinfo {author} {\bibfnamefont {J.~D.}\ \bibnamefont
  {Bryngelson}}, \bibinfo {author} {\bibfnamefont {J.~N.}\ \bibnamefont
  {Onuchic}}, \bibinfo {author} {\bibfnamefont {N.~D.}\ \bibnamefont {Socci}},
  \ and\ \bibinfo {author} {\bibfnamefont {P.~G.}\ \bibnamefont {Wolynes}},\
  }\href@noop {} {\bibfield  {journal} {\bibinfo  {journal} {Proteins}\
  }\textbf {\bibinfo {volume} {21}},\ \bibinfo {pages} {167} (\bibinfo {year}
  {1995})}\BibitemShut {NoStop}%
\bibitem [{Note1()}]{Note1}%
  \BibitemOpen
  \bibinfo {note} {Of course, the paradigm that ground-states correspond to
  free energy minima of folding trajectories implicitly assumes that folding
  does not get stuck in kinetic traps which lead to topologically frustrated
  (unknotted) states. The investigation of this interesting and challenging
  topic is, however, beyond the scope of this paper.}\BibitemShut {Stop}%
\bibitem [{\citenamefont {B{\"o}linger}(2008)}]{boelinger:08}%
  \BibitemOpen
  \bibfield  {author} {\bibinfo {author} {\bibfnamefont {D.}~\bibnamefont
  {B{\"o}linger}},\ }\emph {\bibinfo {title} {Topologische Untersuchungen von
  Proteinen, Homo- und Heteropolymeren}},\ \href@noop {} {Master's thesis},\
  \bibinfo  {school} {Johannes Gutenberg University, Mainz, Germany} (\bibinfo
  {year} {2008})\BibitemShut {NoStop}%
\bibitem [{\citenamefont {Hsu}\ \emph {et~al.}(2003{\natexlab{a}})\citenamefont
  {Hsu}, \citenamefont {Mehra}, \citenamefont {Nadler},\ and\ \citenamefont
  {Grassberger}}]{hsu:phys_rev_E:03}%
  \BibitemOpen
  \bibfield  {author} {\bibinfo {author} {\bibfnamefont {H.-P.}\ \bibnamefont
  {Hsu}}, \bibinfo {author} {\bibfnamefont {V.}~\bibnamefont {Mehra}}, \bibinfo
  {author} {\bibfnamefont {W.}~\bibnamefont {Nadler}}, \ and\ \bibinfo {author}
  {\bibfnamefont {P.}~\bibnamefont {Grassberger}},\ }\href@noop {} {\bibfield
  {journal} {\bibinfo  {journal} {Phys.\ Rev.\ E}\ }\textbf {\bibinfo {volume}
  {68}},\ \bibinfo {pages} {021113} (\bibinfo {year}
  {2003}{\natexlab{a}})}\BibitemShut {NoStop}%
\bibitem [{\citenamefont {Hsu}\ \emph {et~al.}(2003{\natexlab{b}})\citenamefont
  {Hsu}, \citenamefont {Mehra}, \citenamefont {Nadler},\ and\ \citenamefont
  {Grassberger}}]{hsu:j_chem_phys:03}%
  \BibitemOpen
  \bibfield  {author} {\bibinfo {author} {\bibfnamefont {H.-P.}\ \bibnamefont
  {Hsu}}, \bibinfo {author} {\bibfnamefont {V.}~\bibnamefont {Mehra}}, \bibinfo
  {author} {\bibfnamefont {W.}~\bibnamefont {Nadler}}, \ and\ \bibinfo {author}
  {\bibfnamefont {P.}~\bibnamefont {Grassberger}},\ }\href@noop {} {\bibfield
  {journal} {\bibinfo  {journal} {J.\ Chem.\ Phys.}\ }\textbf {\bibinfo
  {volume} {118}},\ \bibinfo {pages} {444} (\bibinfo {year}
  {2003}{\natexlab{b}})}\BibitemShut {NoStop}%
\bibitem [{\citenamefont {Backofen}\ and\ \citenamefont
  {Will}(2006)}]{backofen:constraints:06}%
  \BibitemOpen
  \bibfield  {author} {\bibinfo {author} {\bibfnamefont {R.}~\bibnamefont
  {Backofen}}\ and\ \bibinfo {author} {\bibfnamefont {S.}~\bibnamefont
  {Will}},\ }\href@noop {} {\bibfield  {journal} {\bibinfo  {journal}
  {Constraints}\ }\textbf {\bibinfo {volume} {11}},\ \bibinfo {pages} {5}
  (\bibinfo {year} {2006})}\BibitemShut {NoStop}%
\bibitem [{\citenamefont {Zhang}\ \emph {et~al.}(2007)\citenamefont {Zhang},
  \citenamefont {Kou},\ and\ \citenamefont {Liu}}]{zhang:j_chem_phys:07}%
  \BibitemOpen
  \bibfield  {author} {\bibinfo {author} {\bibfnamefont {J.}~\bibnamefont
  {Zhang}}, \bibinfo {author} {\bibfnamefont {S.~C.}\ \bibnamefont {Kou}}, \
  and\ \bibinfo {author} {\bibfnamefont {J.~S.}\ \bibnamefont {Liu}},\
  }\href@noop {} {\bibfield  {journal} {\bibinfo  {journal} {J.\ Chem.\ Phys.}\
  }\textbf {\bibinfo {volume} {126}},\ \bibinfo {pages} {225101} (\bibinfo
  {year} {2007})}\BibitemShut {NoStop}%
\bibitem [{\citenamefont {W\"ust}\ and\ \citenamefont
  {Landau}(2009)}]{wust:phys_rev_lett:09}%
  \BibitemOpen
  \bibfield  {author} {\bibinfo {author} {\bibfnamefont {T.}~\bibnamefont
  {W\"ust}}\ and\ \bibinfo {author} {\bibfnamefont {D.~P.}\ \bibnamefont
  {Landau}},\ }\href@noop {} {\bibfield  {journal} {\bibinfo  {journal} {Phys.\
  Rev.\ Lett.}\ }\textbf {\bibinfo {volume} {102}},\ \bibinfo {pages} {178101}
  (\bibinfo {year} {2009})}\BibitemShut {NoStop}%
\bibitem [{\citenamefont {W\"ust}\ and\ \citenamefont
  {Landau}(2012)}]{wust:j_chem_phys:12}%
  \BibitemOpen
  \bibfield  {author} {\bibinfo {author} {\bibfnamefont {T.}~\bibnamefont
  {W\"ust}}\ and\ \bibinfo {author} {\bibfnamefont {D.~P.}\ \bibnamefont
  {Landau}},\ }\href@noop {} {\bibfield  {journal} {\bibinfo  {journal} {J.\
  Chem.\ Phys.}\ }\textbf {\bibinfo {volume} {137}},\ \bibinfo {pages} {064903}
  (\bibinfo {year} {2012})}\BibitemShut {NoStop}%
\bibitem [{\citenamefont {Wang}\ and\ \citenamefont
  {Landau}(2001{\natexlab{a}})}]{wang:phys_rev_lett:01}%
  \BibitemOpen
  \bibfield  {author} {\bibinfo {author} {\bibfnamefont {F.}~\bibnamefont
  {Wang}}\ and\ \bibinfo {author} {\bibfnamefont {D.~P.}\ \bibnamefont
  {Landau}},\ }\href@noop {} {\bibfield  {journal} {\bibinfo  {journal} {Phys.\
  Rev.\ Lett.}\ }\textbf {\bibinfo {volume} {86}},\ \bibinfo {pages} {2050}
  (\bibinfo {year} {2001}{\natexlab{a}})}\BibitemShut {NoStop}%
\bibitem [{\citenamefont {Wang}\ and\ \citenamefont
  {Landau}(2001{\natexlab{b}})}]{wang:phys_rev_E:01}%
  \BibitemOpen
  \bibfield  {author} {\bibinfo {author} {\bibfnamefont {F.}~\bibnamefont
  {Wang}}\ and\ \bibinfo {author} {\bibfnamefont {D.~P.}\ \bibnamefont
  {Landau}},\ }\href@noop {} {\bibfield  {journal} {\bibinfo  {journal} {Phys.\
  Rev.\ E}\ }\textbf {\bibinfo {volume} {64}},\ \bibinfo {pages} {056101}
  (\bibinfo {year} {2001}{\natexlab{b}})}\BibitemShut {NoStop}%
\bibitem [{\citenamefont {Lesh}\ \emph {et~al.}(2003)\citenamefont {Lesh},
  \citenamefont {Mitzenmacher},\ and\ \citenamefont
  {Whitesides}}]{lesh:recomb:03}%
  \BibitemOpen
  \bibfield  {author} {\bibinfo {author} {\bibfnamefont {N.}~\bibnamefont
  {Lesh}}, \bibinfo {author} {\bibfnamefont {M.}~\bibnamefont {Mitzenmacher}},
  \ and\ \bibinfo {author} {\bibfnamefont {S.}~\bibnamefont {Whitesides}},\
  }in\ \href@noop {} {\emph {\bibinfo {booktitle} {Proceedings of the 7th
  Annual International Conference on Research in Computational Molecular
  Biology}}}\ (\bibinfo {year} {2003})\ p.\ \bibinfo {pages} {188}\BibitemShut
  {NoStop}%
\bibitem [{\citenamefont {Deutsch}(1997)}]{deutsch:j_chem_phys:97}%
  \BibitemOpen
  \bibfield  {author} {\bibinfo {author} {\bibfnamefont {J.~M.}\ \bibnamefont
  {Deutsch}},\ }\href@noop {} {\bibfield  {journal} {\bibinfo  {journal} {J.\
  Chem.\ Phys.}\ }\textbf {\bibinfo {volume} {106}},\ \bibinfo {pages} {8849}
  (\bibinfo {year} {1997})}\BibitemShut {NoStop}%
\bibitem [{\citenamefont {Millett}\ \emph {et~al.}(2005)\citenamefont
  {Millett}, \citenamefont {Dobay},\ and\ \citenamefont
  {Stasiak}}]{millett:macromolecules:05}%
  \BibitemOpen
  \bibfield  {author} {\bibinfo {author} {\bibfnamefont {K.}~\bibnamefont
  {Millett}}, \bibinfo {author} {\bibfnamefont {A.}~\bibnamefont {Dobay}}, \
  and\ \bibinfo {author} {\bibfnamefont {A.}~\bibnamefont {Stasiak}},\
  }\href@noop {} {\bibfield  {journal} {\bibinfo  {journal} {Macromolecules}\
  }\textbf {\bibinfo {volume} {38}},\ \bibinfo {pages} {601} (\bibinfo {year}
  {2005})}\BibitemShut {NoStop}%
\bibitem [{\citenamefont {Tubiana}\ \emph {et~al.}(2011)\citenamefont
  {Tubiana}, \citenamefont {Orlandini},\ and\ \citenamefont
  {Micheletti}}]{tubiana:prog_theor_phys_suppl:11}%
  \BibitemOpen
  \bibfield  {author} {\bibinfo {author} {\bibfnamefont {L.}~\bibnamefont
  {Tubiana}}, \bibinfo {author} {\bibfnamefont {E.}~\bibnamefont {Orlandini}},
  \ and\ \bibinfo {author} {\bibfnamefont {C.}~\bibnamefont {Micheletti}},\
  }\href@noop {} {\bibfield  {journal} {\bibinfo  {journal} {Prog.\ Theor.\
  Phys.\ Suppl.}\ }\textbf {\bibinfo {volume} {191}},\ \bibinfo {pages} {192}
  (\bibinfo {year} {2011})}\BibitemShut {NoStop}%
\bibitem [{\citenamefont {Millett}\ \emph {et~al.}(2013)\citenamefont
  {Millett}, \citenamefont {Rawdon}, \citenamefont {Stasiak},\ and\
  \citenamefont {Su{\l}kowska}}]{millett:biochem_soc_trans:13}%
  \BibitemOpen
  \bibfield  {author} {\bibinfo {author} {\bibfnamefont {K.~C.}\ \bibnamefont
  {Millett}}, \bibinfo {author} {\bibfnamefont {E.~J.}\ \bibnamefont {Rawdon}},
  \bibinfo {author} {\bibfnamefont {A.}~\bibnamefont {Stasiak}}, \ and\
  \bibinfo {author} {\bibfnamefont {J.~I.}\ \bibnamefont {Su{\l}kowska}},\
  }\href@noop {} {\bibfield  {journal} {\bibinfo  {journal} {Biochem.\ Soc.\
  Trans.}\ }\textbf {\bibinfo {volume} {41}},\ \bibinfo {pages} {533} (\bibinfo
  {year} {2013})}\BibitemShut {NoStop}%
\bibitem [{\citenamefont {Virnau}(2010)}]{virnau:phys_procedia:10}%
  \BibitemOpen
  \bibfield  {author} {\bibinfo {author} {\bibfnamefont {P.}~\bibnamefont
  {Virnau}},\ }\href@noop {} {\bibfield  {journal} {\bibinfo  {journal} {Phys.\
  Procedia}\ }\textbf {\bibinfo {volume} {6}},\ \bibinfo {pages} {117}
  (\bibinfo {year} {2010})}\BibitemShut {NoStop}%
\bibitem [{Note2()}]{Note2}%
  \BibitemOpen
  \bibinfo {note} {Note that for an estimate of the free energy minimum
  knottedness, i.\protect \tmspace +\thinmuskip {.1667em}e., at zero
  temperature, an accurate determination of the density of states is irrelevant
  and WL sampling has merely been used as a powerful Monte Carlo driver to
  sample statistically uncorrelated, low-energy conformations.}\BibitemShut
  {Stop}%
\end{thebibliography}%

\end{document}